\documentstyle[preprint,prl,aps,epsfig]{revtex}
\newcommand \be{\begin{equation}}
\newcommand \bea{\begin{eqnarray}}
\newcommand \ee{\end{equation}}
\newcommand \eea{\end{eqnarray}}

[1999/03/29 PSNFSS v.7.2
Times font as default roman
: S Rahtz]

\begin{document}

\title{A Nonlinear Super-Exponential Rational Model of Speculative Financial Bubbles}

\author{D Sornette$^{1,2}$ and J.V. Andersen$^1$\\
$^1$ Laboratoire de Physique de la Mati\`{e}re Condens\'{e}e\\ CNRS UMR6622 and
Universit\'{e} de Nice-Sophia Antipolis\\ B.P. 71, Parc
Valrose, 06108 Nice Cedex 2, France \\
$^2$ Institute of Geophysics and
Planetary Physics and Department of Earth and Space Science\\ 
University of California, Los Angeles, California 90095\\
e-mails: vitting@unice.fr and sornette@unice.fr\\}

\date{\today}
\maketitle

\vskip 2cm
{\bf Abstract:} 
Keeping a basic tenet of economic theory, rational expectations, we model
the nonlinear positive feedback between agents in the stock market
as an interplay between nonlinearity and multiplicative noise. The derived
hyperbolic stochastic finite-time singularity formula transforms a Gaussian
white noise into a rich time series possessing 
all the stylized facts of empirical prices, as well as 
accelerated speculative bubbles preceding crashes. We use the
formula to invert the two years of price 
history prior to the recent crash on the Nasdaq
(april 2000) and prior to the crash in the Hong Kong market associated
with the Asian crisis in early 1994. These complex price dynamics are 
captured using only one exponent controlling the explosion, the variance and mean of the
underlying random walk. This offers a new and powerful detection tool of
speculative bubbles and herding behavior.

%\narrowtext
%\twocolumn[...text...]

\newpage

\section{Introduction}

Economic structures and financial markets are among the most studied examples
of complex systems \cite{Andersoncomp}, together with
biological and geological networks, which are characterized by the self-organization
of macroscopic ``emergent'' properties. 
One such remarkable behavior is the occurrence of
intermittent accelerated self-reinforcing behavior \cite{PNASmyreview}, such as in the maturation
of the mother-fetus complex culminating in parturition \cite{Parturition1}, 
in the observed accelerated seismicity ending in a great earthquake \cite{SykesJaume,Samsor},
in positive-feedbacks in technology (Betamax versus VHS video standards)
\cite{arthururn2} or in
the herding of speculators preceeding crashes \cite{JSL}. 
The key concept underlying all these systems is the existence of nonlinear
positive feedback.

Here, we formulate a model of such a self-reinforcing behavior in the context
of speculative financial bubbles based on the 
interplay between two key ingredients, multiplicative noise and 
nonlinear positive feedback. In the Stratonovich representation 
usually practiced by physicists, 
our fundamental stochastic dynamical equation for the bubble price $B(t)$ 
is of the form
\be
{d B \over dt} = (a \mu_0 + b \eta)~B^m~,  \label{jfja}
\ee
where $a$ and $b$ are two positive constants and $\eta$ 
is a delta-correlated Gaussian white noise.
The nonlinearity $B^m$ with exponent $m>1$
creates a singularity at some finite time $t^*$
and the multiplicative noise turns out to make $t^*$ stochastic (i.e., dependent
upon the realization of the noise). As we shall show in more details below,
model (\ref{jfja}) with $m=1$ is the standard geometrical Brownian motion
used for describing financial time series at a first-order of approximation.
However, the idea that financial 
time series require inherently nonlinear processes has been firmly established
in the financial literature (see below). But only recently has this idea
been tested in simple models, such as in percolation models of the stock market
\cite{PandeyStauffer}, generalized with two competing nonlinearities in a
dynamical system of price behavior
\cite{SorIde1,SorIde2} and in tests adding different types of noise
\cite{Proykova}. All these works \cite{PandeyStauffer,SorIde1,SorIde2,Proykova}
were aimed specifically at finding what ingredients may cause the 
approximate log-periodic
undulations which have been documented to decorate accelerating 
bubble prices (see \cite{SorJohQF} for
a recent review of the state of the art and references therein).
In the present paper, our goal is to take a step back from the model of 
a speculative bubble in terms of a power law acceleration decorated by a log-periodic
oscillations and ask how a power law acceleration alone together with noise interact
and describe a part (and what part?) of the stylized facts observed in financial
markets. Specifically, we propose in a first step to understand
the interplay between positive nonlinear feedback and 
multiplicative noise. Log-periodicity is an additional characteristics
not captured by our present model. It will be added later when the very
rich phenomenology resulting from model (\ref{jfja}) is fully explored.

In section 2, we put our model in the perspective of the existing research in 
economics and finance, in the goal of showing that it derives in a natural
way from the accumulated evidence and the existing concepts.
In section 3, we present our model and solve it using the formalism of mathematical finance
and Ito calculus (all technical aspects are put in the Appendix), which
provides a controlled definition of the multiplicative noise. In section 4, we 
propose a calibration of the model with two financial time series coming from periods of strong
market acceleration in the Hang Seng index of the Hong Kong 
market prior to the crash which occurred
in early 1994 and in the Nasdaq composite index prior to the crash of April 2000. 
Section 5 concludes.

\section{Previous works on financial bubbles}

According to the efficient market hypothesis,
the movement of financial prices are an immediate
and unbiased reflection of incoming news about future earning prospects.
Thus, any deviation from the random
walk observed empirically would simply reflect similar deviations
in extraneous signals feeding the market. In contrast, a large variety of models
have been developed in the economic, finance and more recently physical literature
which suggest that self-organization of the 
market dynamics is sufficient to create complexity endogenously.
A relatively new school of research, championed in 
particular by the Santa Fe
Institute in New Mexico \cite{Arthuralgogen3,Farmer} 
and being developed now in several
other institutions worldwide \cite{Labmitlo2,BrockHomme5,Lebaronrev,minoritygames1}, 
views markets as
complex evolutionary adaptive systems populated by boundedly rational agents
interacting with each other. 
Several works have modelled the epidemics of opinion and speculative
bubbles in financial markets from an adaptative agent point-of-view
\cite{Kirman,Lux1,LuxMarchesi}. Other 
relevant works put more emphasis on the heterogeneity and threshold
nature of decision making which lead in general to irregular cycles and critical
behavior \cite{Hubermancrash,Taka2,Staubook,Solomon2,Gaunersdorfer}.
Experimental approaches to economics, started
in the the mid-20th century, have also been actively used to examine
propositions implied by economic theories of markets \cite{GilletteDelMas,vernonsmith3}. 
In much of the literature on experimental economics 
\cite{DavisHolt,KagelRoth}, the rational expectations model has been the main
benchmark against which to check the informational efficiency of experimental markets. 
Experiments on markets with insiders and uninformed traders \cite{PlottSunder2} 
show that equilibrium prices do reveal insider information 
after several trials of the experiments, suggesting that the markets
disseminate information efficiently, albeit under restricted conditions 
\cite{PlottSunder2,Forsythe}.

Notwithstanding a plethora of models which 
account approximately for the main stylized facts observed in stock markets, the
characteristic structure of speculative bubbles is not captured at all.
However, if speculative bubbles do exist, they probably constitute one of the most
important empirical fact to explain and predict, due to their psychological effects
(as witnessed by the medias and popular as well as economic press) 
and their financial impacts (potential losses of up to
trillions of dollars during crashes and recession following these bubbles).
Since the publication of the original contributions on rational
expectations (RE) bubbles \cite{Blanchard,Blanwat}, a
large literature has indeed emerged on theoretical refinements of the original concept
and on the empirical detectability of RE bubbles in financial data (see 
\cite{Camerer} and \cite{Adam} for surveys of this literature). Empirical
research has largely concentrated on testing for explosive exponential trends in the time
series of asset prices and foreign exchange rates \cite{Evans,Woo}, with
however limited success. The first reason lies
in the absence of a general definition, as bubbles are model specific and generally
defined from a rather restrictive framework.  The concept of a fundamental
price reference does not necessarily exist, nor is it necessarily unique. Many RE
bubbles exhibit shapes that are hard to reconcile with the economic intuition
or facts \cite{Luxsornette}. 
A major problem is that apparent evidence for
bubbles can be reinterpreted in terms of market fundamentals that are
unobserved by the researcher. Another suggestion is that, if
stock prices are not more explosive than dividends, then it can be
concluded that rational bubbles are not present, since bubbles are
taken to generate an explosive component to stock prices.  However, 
periodically collapsing bubbles are not detectable by using
standard tests to determine whether stock prices are more explosive or
less stationary than dividends \cite{Evans}. In sum, the present evidence
for speculative bubbles is fuzzy and unresolved at best.

\section{The positive feedback model with multiplicative noise}

Keeping a basic tenet of economic theory, rational expectations, we model
the nonlinear positive feedback between agents
as an interplay between nonlinearity and multiplicative noise. The derived
hyperbolic stochastic finite-time singularity formula transforms a Gaussian
white noise into a rich time series possessing 
all the stylized facts of empirical prices, i.e.,
no correlation of returns \cite{Campbelletal}, 
long-range correlation of volatilities \cite{Dingetal}, fat-tail
of return distributions \cite{Mandelbrot1,devries,MantegnaStan}, 
apparent multifractality \cite{Mandel97,Muzyb}, sharp peak-flat trough
pattern of price peaks \cite{RoehnerSorSP} as well as 
accelerated speculative bubbles preceding crashes \cite{JSL}. 

The most important feature of our model is that bubbles 
are growing ``super-exponentially'', 
i.e., with a growth growing itself with time leading to a power law acceleration
leading in principle to a singularity.
Our super-exponential bubbles are thus fundamentally different from
all previous bubble models based on exponential growth (with constant
average growth rate). This novel property provides a much clearer
procedure for testing for the presence of bubbles in empirical data: rather
than trying to detect an anomalous exponential growth as performed
in essentially all previous tests which is easily confused with the ``normal''
behavior of the fundamental price, we propose that the super-exponential 
growth of bubbles provides a clear distinguishing signature. 

We start from the celebrated geometric Brownian model of the 
bubble price $B(t)$,
solidified into a paradigm by Black-Scholes option pricing model \cite{Merton},
$d B = \mu B dt + \sigma B dW_t$,
where $\mu$ is the instantaneous return rate, $\sigma$ is the volatility
and $dW_t$ is the infinitesimal increment of the random walk with unit variance
(Wiener process). 
We generalize this expression into
\be
d B(t) = \mu(B(t)) B(t) dt + \sigma(B(t)) B(t) dW_t - \kappa(t) B(t) dj~,  \label{nfkaaak}
\ee
allowing $\mu(B(t))$ and $\sigma(B(t))$ to depend arbitrarily 
and nonlinearly on the instantaneous realization of the price.
A jump term has been added to describe a correction or a crash
of return amplitude $\kappa$, which can be a stochastic variable taken from 
an a priori arbitrary distribution. 
Immediately after the last crash which becomes the new origin of time $0$,
$dj$ is reset to $0$ and will eventually jump to $1$ with a
hazard rate $h(t)$, defined such that the probability that a crash occurs between $t$ 
and $t+dt$ conditioned
on not having occurred since time $0$ is $h(t) dt$. Here, we follow 
well-established models 
Cox, Ross and Rubinstein \cite{cosrosru} and Merton \cite{Merton76}
to define the jump $dj$ as a {\it discontinuous} process. Specifically,
conditioned on the fact that the jump has not occurred until time $t$,
in the next time increment $dt$, the jump from $0$ to $1$ occurs with probability
$h(t)dt$ and does not occur with probability $1-h(t)dt$. Hence, its average
$\langle dj \rangle$ is 
\be
\langle dj \rangle = 1 \times h(t) dt + 0 \times (1-h(t)dt) = h(t) dt~.
\ee

Following \cite{Blanchard,Blanwat}, $B(t)$ is a 
rational expectations bubble which accounts for the possibility, often
discussed in the empirical literature and by practitioners, that observed
prices may deviate significantly and over extended time intervals
from fundamental prices.
While allowing for deviations from fundamental prices, rational bubbles
keep a fundamental anchor point of economic modelling, namely that bubbles
must obey the condition of rational expectations. This translates essentially
into the no-arbitrage condition with risk-neutrality, which states that
the expectation of $d B(t)$ conditioned on the past
up to time $t$ is zero. This allows us to determine 
the crash hazard rate $h(t)$ as a function of $B(t)$.
Using the definition of the hazard rate $h(t) dt = \langle dj \rangle$,
where the bracket denotes the expectation over all possible outcomes since the last crash, 
this leads to 
\be
\mu(B(t)) B(t)   - \langle \kappa \rangle B(t) h(t) = 0~,
\ee
which provides the variable hazard rate:
\be
h(t) = {\mu(B(t)) \over \langle \kappa \rangle}~.  \label{bvfjuaj}
\ee
Expression (\ref{bvfjuaj}) quantifies the fact that the theory of rational expectations
with risk-neutrality
associates a risk to any price: for example, if the bubble price explodes, so will the crash
hazard rate, so that the risk-return trade-off is always obeyed. 
This model generalizes Refs.~\cite{JSL,JLS} by driving the hazard rate by the price,
rather than the reverse. 

However, most investors are risk-averse rather than risk-neutral and this
risk aversion is likely to be crucial in extreme situations, such as preceding crashes.
As already discussed \cite{JSL}, there are two ways
of incorporating risk aversion. The first one consists
in introducing a risk premium rate $r_R \in (0,1]$ such that the 
no-arbitrage condition on the bubble price reads
$(1-r_Rdt) {\rm E}_t[B(t+dt)]= B(t)$, 
where ${\rm E}_t[y]$ denotes the expectation of $y$ conditioned on 
the whole past history up to time $t$. Putting (\ref{nfkaaak}) into this condition
recovers (\ref{bvfjuaj}) with $\mu(B(t))$ changed into $\mu(B(t))-r_R$: for
a given market return $\mu(B(t))$, risk aversion implies a smaller
crash hazard rate; reciprocally, a given crash hazard rate requires a large
market return in the presence of risk aversion. The
introduction of the risk aversion rate $r_R$ has only the effect of redefining
the effective market return and does not change the results presented below.
In particular, $r_R=\eta \mu(B(t))$ captures the fact that the
risk-premium that risk-averse investors demand increases with the level of risk.
This specification amounts simply to change $\mu(B(t))$ by $(1-\eta)\mu(B(t))$
in the following and does not modify either our main conclusions.

Another way to incorporate risk aversion is to say that
the probability of a crash in the next instant is perceived by traders
as being $K$ times bigger than it objectively is. This amounts to
multiplying the crash hazard rate $h(t)$ by $K$ and therefore does not
either modify the structure of $h(t)$.
The coefficients $r_R$ and $K$ both represent general aversion
of fixed magnitude against risks.  Risk aversion is a central feature
of economic theory and is generally thought to be stable within a
reasonable range being associated with slow-moving secular trends like
changes in education, social structures and technology.
Risk perceptions are however constantly changing in the course of real-life 
bubbles. This is indeed captures by our model in which risk
perceptions quantified by $h(t)$ do oscillate dramatically throughout 
the bubble, even though subjective aversion to risk remains stable, simply because the
objective degree of risk that the bubble may burst goes through
wild swings.

We now specify the dependence of $\mu(B(t))$ and $\sigma(B(t))$
to capture the possible appearance of
positive feedbacks on prices. There are many mechanisms in the stock
market and in the behavior of investors which may lead to positive feedbacks.
First, investment strategies with ``portfolio insurance'' are such that
sell orders are issued whenever a loss threshold (or stop loss) is passed. It is clear that
by increasing the volume of sell order, this may 
lead to further price decreases.  Some commentators like Kim and Markowitz 
\cite{KimMarkowitz} have indeed attributed the 
crash of Oct. 1987 to a cascade of sell orders.
Second, there is a growing empirical evidence of the existence of herd or ``crowd'' 
behavior in speculative markets \cite{Shillerexu}, 
in fund behaviors \cite{Schar,Grinblatt}  and 
in the forecasts made by financial analysts \cite{Trueman}. 
Although this behavior is inefficient from a social
standpoint, it can be rational from the perspective of managers who are
concerned about their reputations in the labor market.
Such behavior can be rational
and may occur as an information cascade, a situation in which every subsequent actor, 
based on the observations of others, makes the same choice independent of
his/her private signal \cite{Welch3}. 
Herding leads to positive nonlinear feedback. Another mechanism for positive
feedbacks is the so-called ``wealth'' effect: a rise of the stock market
increases the wealth of investors who spend more, adding to the earnings
of companies, and thus increasing the value of their stock.

The evidence for nonlinearity has a strong empirical support: for instance, 
the coexistence of
the absence of correlation of price changes and the strong autocorrelation of their 
absolute values can not be explained by any linear model \cite{Hsieh2}.
Comparing additively nonlinear processes and multiplicatively nonlinear 
models, the later class of models are found consistent
with empirical price changes and with options' implied volatilities \cite{Hsieh2}.
With the additional insight that hedging strategies of general
Black-Scholes option models lead to a positive feedback on the volatility \cite{Sircar},
we are led to propose a nonlinear model with multiplicative noise in which
the return rate and the 
volatility are nonlinear increasing power law of $B(t)$: 
\bea
\mu(B) B &=& {m \over 2B} [B \sigma(B)]^2 + \mu_0 [B(t)/B_0]^m ~,  \label{buyaauqka}   \\
\sigma(B) B &=&  \sigma_0 [B(t)/B_0]^m~,  \label{fjallqaaq}
\eea
where $B_0$, $\mu_0$, $m>0$ and $\sigma_0$ are four parameters of the model, setting respectively
a reference scale, an effective drift and the strength of the nonlinear 
positive feedback. The first term in the r.h.s. (\ref{buyaauqka}) is added as 
a convenient device to simplify the Ito calculation of these stochastic differential
equations. The model can be reformulated in the Stratonovich interpretation
given by expression (\ref{jfja}).
Recall that, in physicist's notation, $\eta dt \equiv dW$. The form (\ref{jfja})
examplifies the fundamental ingredient of our theory
based on the interplay between nonlinearity and multiplicative noise. The nonlinearity
creates a singularity in finite time 
and the multiplicative noise makes it stochastic. The choice (\ref{buyaauqka},\ref{fjallqaaq})
or (\ref{jfja}) are the simplest generalisation of the standard geometric Brownian model
(\ref{nfkaaak}) recovered for the special case $m=1$. The introduction of the exponent $m$
is a straightforward mathematical trick to account in the simplest and most parsimonious
way for the presence of nonlinearity. Note in particular that, in the limit where $m$ becomes
very large, the nonlinear function $B^m$ tends to a threshold response. The power
$B^m$ can be decomposed as $B^m = B^{m-1} \times B$ stressing the fact that $B^{m-1}$ plays
the role of a growth rate, function of the price itself. The positive feedback effect
is captured by the fact that a larger price $B$ feeds a larger growth rate, which leads
to a larger price and so no. We do not attempt to unravel the specific mechanisms behind
herding and positive feedbacks, rather we model this behavior in the simplest possible
mathematical manner.

The
solution of (\ref{nfkaaak}) with (\ref{buyaauqka}) and (\ref{fjallqaaq})
is derived in the Appendix (see also \cite{secondpaper} for details of the derivation). 
The bubble price $B(t)$,
conditioned on no crash occuring ($dj=0$), is given by \cite{secondpaper}
\be
B(t) = \alpha^{\alpha}~{1 \over \left(\mu_0[t_c - t] - 
{\sigma_0 \over B_0^{m}}~W(t)\right)^{\alpha}}~,  ~~~~{\rm where}~\alpha\equiv {1 \over m-1}
\label{jfkaaakaaa}
\ee
with $t_c= y_0/(m-1)\mu_0$ is a constant determined by the initial condition
with $y_0=1/B(t=0)^{m-1}$ (see the appendix).
Expression (\ref{jfkaaakaaa}) is our main formal result
and is illustrated in figure 1. To grasp its meaning, let
us first consider the deterministic case $\sigma_0=0$, such that the 
return rate $\mu(B) \propto [B(t)]^{m-1}$ is the sole driving term. Then,
(\ref{jfkaaakaaa}) reduces to $B(t) \propto 1/[t_c-t]^{1 \over m-1}$, i.e.,
a positive feedback $m>1$ of the price $B(t)$ on the return rate $\mu$ 
creates a finite-time singularity at a critical time $t_c$ determined by the initial
starting point. This power law
acceleration of the price accounts for the effect of herding resulting from
the positive feedback. It is in agreement with the empirical finding that
price peaks have sharp concave upwards maxima \cite{RoehnerSorSP}.
Reintroducing the stochastic component $\sigma_0 \neq 0$, we see
from (\ref{jfkaaakaaa}) that the finite-time singularity still exists but
its visit is controlled by the first passage of a biased random walk at the 
position $\mu_0 t_c$ such that the denominator
$\mu_0[t_c - t] -  {\sigma_0 \over B_0^{m}}~W(t)$ vanishes.
In practice, a price trajectory will never sample the finite-time
singularity as it is not allowed to approach too close to it due to the jump process
$dj$ defined in (\ref{nfkaaak}). Indeed, from the no-arbitrage condition, the
expression (\ref{bvfjuaj}) for the crash hazard rate ensures that when the price
explodes, so does $h(t)$ so that a crash will occur with larger and larger probability,
ultimately screening the divergence which can never be reached. The endogeneous 
determination (\ref{bvfjuaj}) of the crash probability also ensures that the denominator 
$\mu_0[t_c - t] - {\sigma_0 \over B_0^{m}}~W(t)$ never becomes negative: when it
approaches zero, $B(t)$ blows up and the crash hazard rate increases accordingly.
A crash will occur with probability $1$ before the denominator reaches zero.
Hence, the price $B(t)$ remains always positive and real.
We stress the
remarkably simple and elegant constraint on the dynamics provided by the 
rational expectation condition
that ensures the existence and stationarity of the dynamics at all times, nothwithstanding
the locally nonlinear stochastic explosive dynamics.
When $\mu_0 >0$, the random walk has a positive drift attracting the denominator 
in (\ref{jfkaaakaaa} to zero
(i.e., attracting the bubble to infinity). However, by the mechanism explained above, 
as $B(t)$ increases, so does the crash hazard rate by the relation (\ref{bvfjuaj}). 
Eventually, a crash occurs that reset the bubble to a lower price. The random walk with drift
goes on, eventually $B(t)$ increases again and reaches ``dangerous watersÕÕ, a crash occurs
again, and so on. Note that a crash is not a certain event: an inflated bubble price can
also deflate spontaneously by the random realisation of the random walk $W(t)$ which 
brings back the denominator far from zero.

\section{Properties of the positive feedback model with multiplicative noise}

From a mathematical point of view, the process
(\ref{nfkaaak}) with (\ref{bvfjuaj}, \ref{buyaauqka}, \ref{fjallqaaq}) exists
only for a finite time, whose duration is a random variable: as can be
seen from the solution (\ref{jfkaaakaaa}), the denominator
$D \equiv \mu_0[t_c - t] -  {\sigma_0 \over B_0^{m}}~W(t)$ is positive at the initial time
$t=0$ (with $W(t=0)=0$) and drifts towards zero with average velocity $\mu_0$
decorated by the random walk $W(t)$.  It is well-known that 
the denominator $D$ goes to zero with probability $1$
and the probability that $D$ remains strictly positive up to time $t$ decays
exponentially fast as $t$ increases, with an algebraic power law prefactor
$1/t^{3/2}$ characteristic of the distribution of first returns to the origin
of a random walk. The leading exponential decay is itself due to the non-zero drift $m_0$
and would disappear in the absence of bias.
Thus, the process(\ref{nfkaaak}) with (\ref{bvfjuaj}, \ref{buyaauqka}, \ref{fjallqaaq}) exists
over finite lifetimes which are exponentially distributed.
The explicit analytical solution (\ref{jfkaaakaaa}) shows that it is unique.
From a finance mathematical point of view, we stress that this model is free of
arbitrage, a property resulting from the introduction of the crash-jump process
with hazard rate $h(t)$ defined by (\ref{bvfjuaj}). However, it is clear
that the market is ``incomplete'' in the technical sense of option/derivative theory
in Finance, as it is not possible to replicate in continuous time
an arbitrary option \cite{Merton} by a portfolio made of the stock obeying the process
(\ref{nfkaaak}) and of a risk-free asset. This is due to the existence of the jump/crash
process: this feature is well-known for jump processes \cite{Merton}.

In agreement with empirical observations, returns $\ln [B(t+\tau)/B(t)]$ are
uncorrelated by definition of the RE dynamics (\ref{nfkaaak}) with (\ref{bvfjuaj}).
The absolute values of the returns
exhibit long-range correlations in good agreement with empirical data. 
Figure 2 shows, as a function of time-lag, the correlation
function of the absolute values of the returns constructed from the process (\ref{fundsum})
taking into account that the observed price is the sum of a fundamental price
$F$ and of the bubble component. The correlation function
decays extremely slowly as a function of time-lag 
with a decay approximately linear in the logarithm
of time \cite{Muzyb} (which is also compatible with a power law decay with a small exponent).
This behavior is associated with clustering of volatility driven by the nonlinear
hyperbolic structure of the dynamics (\ref{jfkaaakaaa}). This result
is obtained by averaging over many bubbles. 
Conditioned on a single bubble and provided no crash has yet occurred,
the correlation function
can actually be non-stationary and grow with time as the bubble approaches 
stochastically the critical
time $t_c$. This prediction of our model is actually born out by measurements of price
dynamics prior to the major crashes, which will be reported in full details 
elsewhere \cite{secondpaper}.

Figure 3 shows that the empirical distribution of returns is also recovered with
no adjustment of parameters. To construct a meaningful distribution, we have 
added a constant fundamental price $F$ to the bubble price $B(t)$ as only their sum is
observable in real life:
\be
P(t) = e^{rt} \left[ F + B(t) \right]~.   \label{fundsum}
\ee
We can also include the possibility for a interest rate $r$ or growth of the economy with
rate $r$.
Different curves for various values of $F$ demonstrate
the remarkable robustness of the distributions with respect to the choice of the
unknown fundamental value.
We observe an approximate power law decay with exponent
close to $1.5$ in an intermediate regime, followed by a faster decay with exponent 
approximately $4$, in agreement with previously reported values 
\cite{MantegnaStan,Gopikrishnan}. These apparent power law result from the
superposition of the contribution of many bubbles approaching towards their finite-time
singularity within varying distances constrained by the 
underlying random walk process and the crash hazard rate. For each single bubble, there is
an exact asymptotic truncated power law behavior that can be obtained analytically
\cite{secondpaper} from the expression (\ref{jfkaaakaaa}). In particular, one
can show that the distribution of return over a complete trajectory of a given 
bubble is a power law with exponent $(m-1)/m$ less than $1$. It is the 
combination of these truncated power laws and the mixture of bubble 
and fundamental prices that give rise to
distributions in agreement with empirical facts. 
This suggests that the attention given to the distribution of returns
in the physical literature may have overemphasized its significance.

To demonstrate that our model captures most of the detailed structure of price
time series at times preceding crashes, we use expression (\ref{fundsum})
with (\ref{jfkaaakaaa}) to invert the real price time series and obtain the value
of the key variables of the model. We focus here on two 
examples, the Hang Seng index of the Hong Kong market prior to the crash which occurred
in early 1994 and the Nasdaq composite index prior to the crash of April 2000. 
Other examples are reported in \cite{secondpaper}. 
To implement the inversion of (\ref{fundsum})
with (\ref{jfkaaakaaa}), we note that if these equations represent 
the market behavior faithfully, then starting from a real price time series $P(t)$,
the times series 
\be
{\hat W}(t) \equiv \left[ P(t) e^{-rt} - F \right]^{-(m-1)}    \label{bggnlas}
\ee
should be a bias random walk, characterized by a constant drift $M=\mu_0/\alpha$
 and volatility
$\sqrt{V}=\sigma_0 / \alpha B_0^{m}$. In other words, the inversion 
(\ref{bggnlas}) should whiten and gaussianize the empirical price series. This 
inversion has the important advantage of not requiring
the determination of the critical time $t_c$ which
appears as a constant term in ${\hat W}(t)$.

To test this hypothesis, we start from an 
arbitrary set of the five parameters $m, V, M, r, F$ of the model and construct
${\hat W}(t)$ using (\ref{bggnlas}). We then analyze ${\hat W}(t)$ to check 
whether it is indeed a pure random walk. For this, we use a battery of tests.
First, we check that the correlation function of the increments $dW$ of $W(t)$
is zero up to the statistical noise. As a second test, we investigate
the distance of the distribution of ${\hat W}(t)$
from a Gaussian distribution. 
We have used the Kolmogorov-Smirnov (KS) test and Anderson-Darling test to qualify
the quality of the Gaussian description of ${\hat W}(t)$. We use the KS distance
as a cost function to minimize to get the optimal set of parameters $m, V, M, r, F$.
We have organized hierarchically the search and find \cite{secondpaper} that 
the two leading parameters explaining most of the data are the exponent $m$ and the
variance $V$ of the random walk as it should.
The quality of the inversion is weakly sensitive to $M$, even less to $r$ and 
almost insensitive to the fundamental value $F$, suggesting that observed prices
at times of 
accelerated bubbles are mostly determined by the bubble component. 
Figure 4 shows the cumulative distributions of the increments $dW$
of the best reconstructed ${\hat W}(t)$
and of the empirical price variations 
and their fit by a cumulative Gaussian distribution, for the Hang Seng 1994 and Nasdaq
2000 bubbles. The inversion procedure is almost perfect for the Hang Seng index
and of good quality but not perfect for the Nasdaq index. For the Hang Seng bubble,
the KS confidence level that the distribution is Gaussian goes from $11\%$ to $96\%$ 
when going from the empirical price to the transformed variable ${\hat W}(t)$
defined by (\ref{bggnlas}). In other words, the whitening inversion is such that
it is not possible to reject the hypothesis that ${\hat W}(t)$ is a genuine random walk,
while the corresponding hypothesis for the empirical price is rejected.
For the Nasdaq bubble, the gain in statistical significance is less striking, from 
$73\%$ to $86\%$ but the visual appearance of the fits is significantly better.

Figure 5 shows ten random time evolutions of the process (\ref{fundsum}) with 
the above best parameter values with distinct random realizations of synthetic
random walks ${\hat W}(t)$ for both bubbles and compare them with the empirical prices.
This figure illustrates the fact that the empirical prices can be seen 
as specific realizations among an ensemble of possible scenarios.
Our model is able to capture remarkably well the visual acceleration of
these indices as a function of time. We stress that standard models of exponential growth
would not give such a good fit.

Our nonlinear model with positive feedback together with the
inversion procedure (\ref{bggnlas}) provides a new direct tool for detecting
bubbles, for identifying their starting times and the plausible ends.
Changing the initial time of the time series, the KS probability of the resulting
Gaussian fit of the transformed series ${\hat W}(t)$ allows us to determine
the starting date beyond which the model becomes inadequate at a
given statistical level. Furthermore, the exponent
$m$ (or equivalently $\alpha$) provides a direct measure of the 
speculative mood. $m=1$ is the normal regime, while $m>1$ quantifies a positive
self-reinforcing feedback. This opens the possibility to continuously monitor
it via the inversion formula (\ref{bggnlas}) and use it
as a ``thermometer'' of speculation, as will be reported elsewhere
\cite{secondpaper}.  Furthermore,
the variance $V$ of the multiplicative noise is a robust measure of volatility.
Its continuous monitoring via the inversion formula (\ref{bggnlas}) suggests
new ways at looking at dependence between assets, in the spirit of but
generalizing the nonlinear transformation of \cite{portfolioreport}. 
Expression (\ref{jfkaaakaaa}) also rationalizes why it is so difficult
to develop reliable statistics on bubbles. Since their occurrence is 
associated with the approach of the random walk $W(t)$ to a level at
which a singularity occurs,
the theory of first approach or of first returns of random walks indicate that the distribution
of waiting times between bubbles has a long tail decaying as $t^{-3/2}$
such that the average waiting time is infinite: one expects to
wait a very long time before observing a bubble following the last one.  
We have indeed verified directly with the numerical simulations that the
distribution of waiting times between consecutive bubbles is the power
law $t^{-3/2}$. This 
feature is a direct prediction of our theory. We note that our theory also
applies to ``anti-bubbles'' or strong ``bear'' regimes, such as the 
behavior of the Nasdaq Composite index since its crash until present times. Positive feedback
can also work to make things worse, not only to hype prices up. 
This will be reported elsewhere \cite{secondpaper}.

\section{Conclusion}

In summary, we have presented a nonlinear model with
positive feedback and multiplicative noise, which explains in a parsimonious
and economically intuitive way
essentially all the characteristics of empirical financial time series, including
the spontaneous emergence of speculative bubbles. It could provide a 
simple starting point for multivariate modelling of financial and economic
variables. We shall report elsewhere the results of our tests using this model
to identify periods of non-linear bubbles from periods of ``normal'' times and
how our model allows us to distinguish these two regimes quantitatively.

\vskip 0.5cm
Acknowledgement: We thank T. Lux for discussions. J.V.A. acknowledges
support from CNRS, France. D.S. gratefully acknowledges
support from the James S. McDonnell Foundation 21st Century Scientist award/studying
complex systems.

\pagebreak
\section*{Appendix: Derivation of the bubble solution}

In this appendix, we derive the solution (\ref{jfkaaakaaa}).
Changing variable from $B$ to $y=\phi(B)=1/B^{m-1}$, Ito calculus tells us that
the coefficients $\mu(B) B$ and $\sigma(B) B$
of an equation of the form $dB= \mu(B) B dt + \sigma(B) B dW$ are changed into
\bea
{\hat \mu}(y) = \mu(B) B {d\phi \over dB} + {1 \over 2} [\sigma(B) B]^2 ~{d^2\phi \over dB^2}
&=& - \mu(B) B {m-1 \over B^m} + {1 \over 2} [\sigma(B) B]^2 {m (m-1) \over B^{m+1}}  \\
{\hat \sigma}(y) = \sigma(B) B {d\phi \over dB} = - \sigma(B) B {m-1 \over B^m}~,
\eea
where $dy = {\hat \mu}(y) dt + {\hat \sigma}(y) dW$.

With the parameterization (\ref{buyaauqka},\ref{fjallqaaq}), 
the equation on $y$ becomes
\be
dy = -(m-1)\mu_0 dt  - {1 \over B_1^{m-1}}~ dW~, \label{gakfala}
\ee
where 
\be
B_1=\left({B_0^m \over (m-1)\sigma_0}\right)^{1/(m-1)}~.
\ee
By the nonlinear change of variable $y=1/B^{m-1}$, we thus recover a simple Brownian
motion with constant drift and constant volatility in the variable $y$.
The solution of (\ref{gakfala}) is 
\be
y(t) = y_0 - (m-1) \mu_0 t - {1 \over B_1^{m-1}}~W(t)~.
\ee
$y_0$ is the initial value $y(0)=y_0=1/B(t=0)^{m-1}$.

In terms of the price $B(t)$, we get (\ref{jfkaaakaaa}) by inverting $y=1/B^{m-1}$
where $t_c= y_0/(m-1)\mu_0$ is a constant determined by the initial condition.
It is important to stress that a finite-time singularity occurs when the denominator
of the right-hand-side of (\ref{jfkaaakaaa}) goes to zero. In absence of noise 
($\sigma_0=0$), $t_c$ is the critical time. However, in the general case with
$\sigma_0>0$, the finite-time singularity occurs at a random time no more equal to $t_c$
which depends on the specific realization of the random walk $W(t)$.

\pagebreak

\pagebreak

\begin{figure}
\begin{center}
\epsfig{file=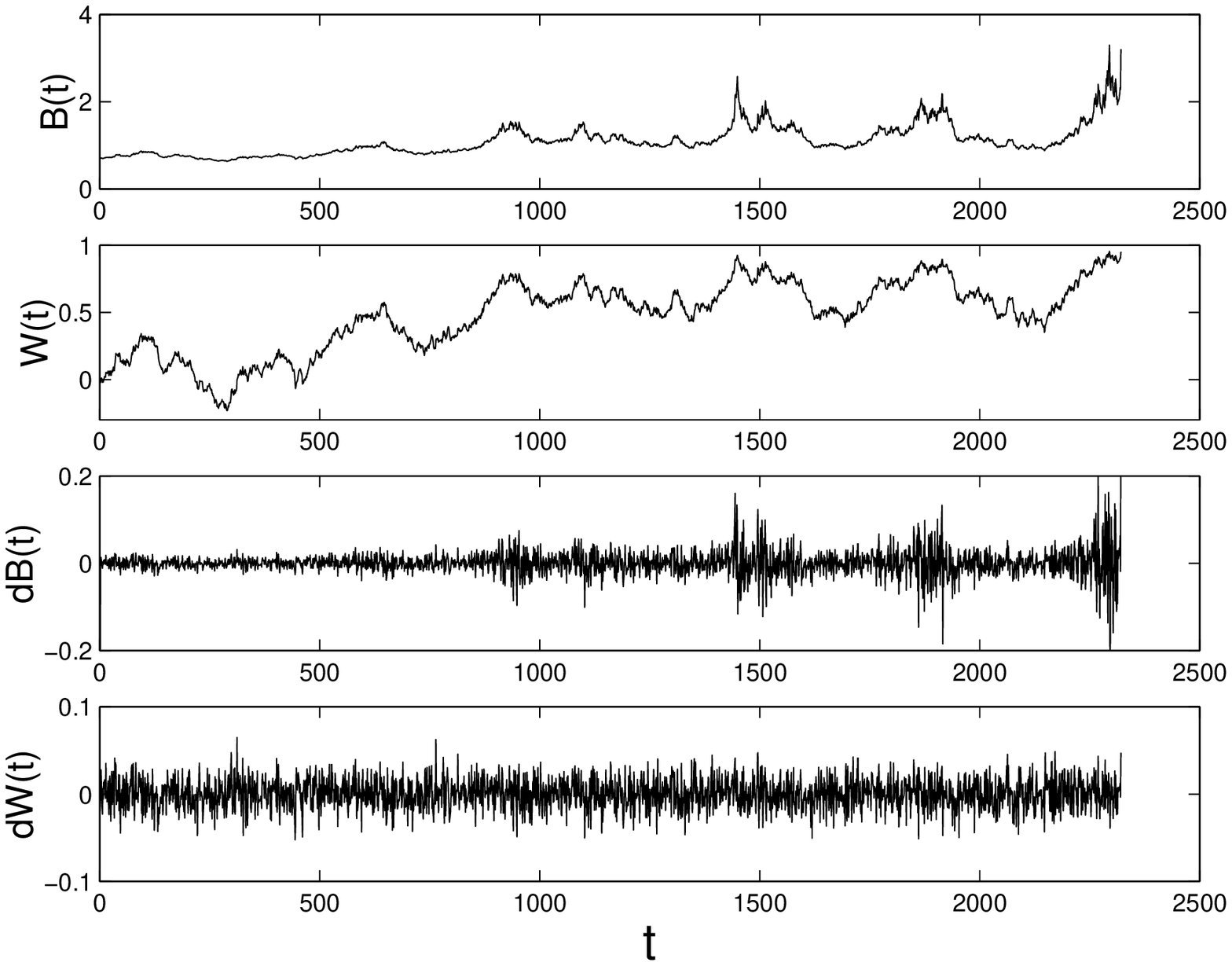,height=8cm,width=12cm}
\caption{\protect\label{Fig1} Typical realization of a bubble (top panel) for 
the parameters $m=3$, $y_0=1$, $\sigma_0=\sqrt{0.0003}$, $\delta t = 3 \cdot 10^{-3}$
(such that one time step corresponds typically to one day of trading on 
the Nasdaq composite index,
calibrated by comparing the daily volatilities), $t_c=1$ and $B_0=1$.
The underlying random walk $W(t)$
(second panel), the bubble daily increments $dB$ (third panel) and random walk increments
$d W$ (bottom panel) are also shown.
Notice the intermittent bursts of strong volatility in the bubble compared to the 
featureless constant level of fluctuations of the random walk.
A numerical simulation of this process requires 
a discretization of the time in steps on size $\delta t$. 
Then, knowing the value of the randow walk 
$W(t-\delta t)$ and the bubble price $B(t-\delta t)$ at the previous time $t-\delta t$, 
we construct $W(t)$ by adding an increment
taken from the centered Gaussian distribution with variance $\delta t$.
From this, we construct $B(t)$ using (\protect\ref{jfkaaakaaa}). We then read off from
(\protect\ref{bvfjuaj}) what is the probability $h(t) \delta t$
for a crash to occur during the next time step.
We compare this probability to a random number $ran$ uniformely drawn in the interval $[0,1]$
and trigger a crash if $ran \leq h(t) \delta t$. In this case, the price $B(t)$ is changed
into $B(t) (1-\kappa)$, where $\kappa$ is drawn from a pre-chosen distribution. In the
simulations presented below, the drop $\kappa$ is fixed at $20\%$. It is straightforward
to generalize to an arbitrary distribution of jumps. After the crash, the dynamics proceeds
incrementally as before, starting from this new value.
If $ran > h(t) \delta t$, no crash occurs and the dynamics can be iterated another time step.
In the time series shown here, there are no crashes, except for the end point. 
We show just one bubble that finally crashes at the end.
 The highly nonlinear formula
(\protect\ref{jfkaaakaaa}) transforms a featureless random walk (second
and fourth panels) into a structured time series
with intermittent volatility bursts (first and third panels).
}
\end{center}
\end{figure}

\newpage

\begin{figure}
\begin{center}
\epsfig{file=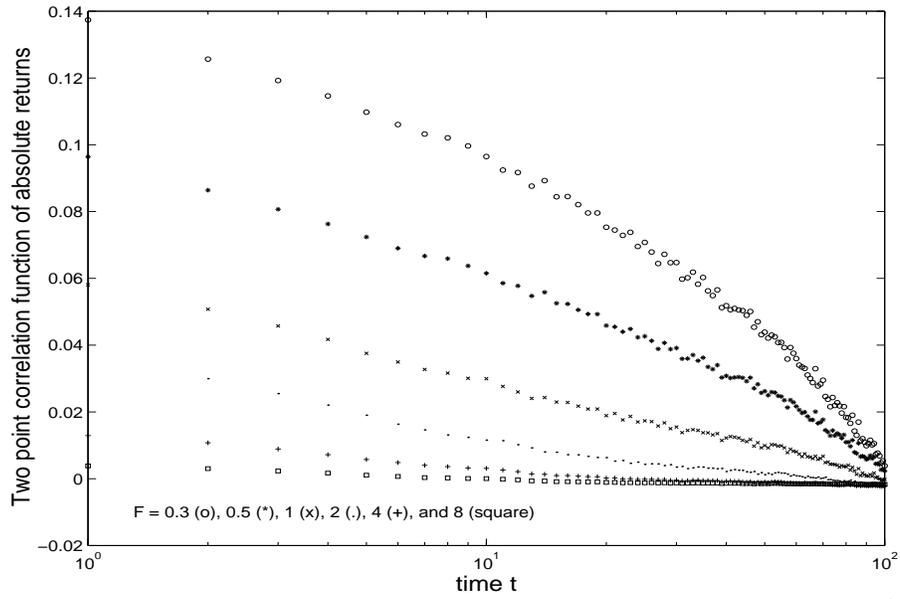,height=8cm,width=12cm}
\caption{\protect\label{Fig2} Two point correlation function of the absolute value of the return of
the price $P(t)$  defined by eq.~(\ref{fundsum}) as a function of time lag
in logarithmic scale (one time step corresponds approximately to one trading day).
The correlation function is calculated as a statistical average over
$300$ independent bubbles B(t), where each bubble was run for $1000$ time steps.
Different points correspond to different values of the
fundamental price $F$. The parameters of the bubbles $B(t)$ are
$m=3, V=0.0003,  \mu_0=0.01,  \kappa=0.2, \sigma_0=\sqrt{0.0003}$ and $B_0=1.0$.
}
\end{center}
\end{figure}

\newpage

\begin{figure}
\begin{center}
\epsfig{file=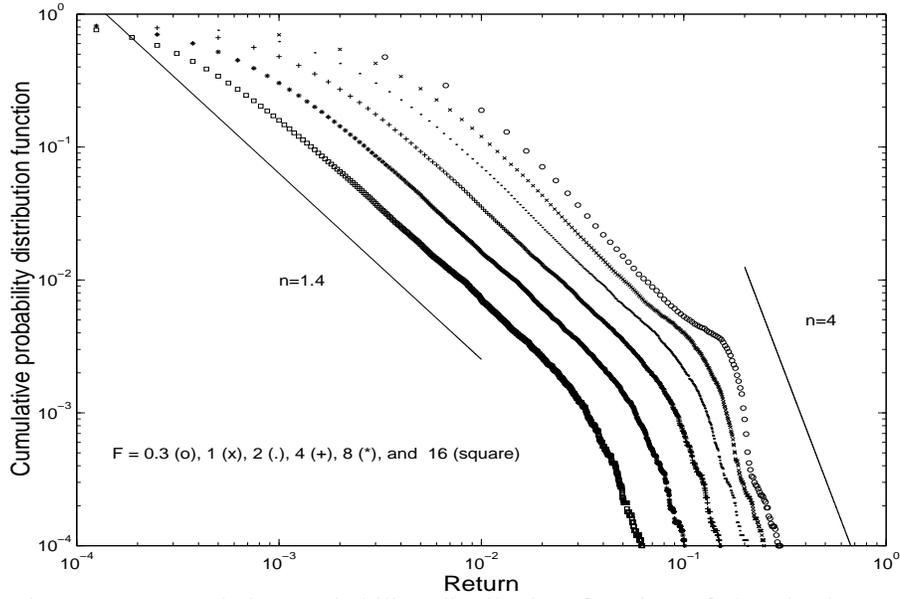,height=8cm,width=12cm}
\caption{\protect\label{Fig3} Complementary cumulative probability distribution function of the
absolute values of the
returns $\ln [dP(t)=P(t)/P(t-1)]$ where $P(t)$ is defined by eq.~(\ref{fundsum}). 
The distribution is symmetric to a good approximation and we thus superimpose the
tails for positive and negative returns.
The probability distribution function is calculated
as an statistical average over 300 independent
bubbles $B(t)$, where each bubble was run for 1000 time steps.
Different points correspond to different values of the
fundamental price $F$. The parameters of the bubbles $B(t)$ are
$m=3, V=0.0003,  \mu_0=0.01,  \kappa=0.2, \sigma_0=\sqrt{0.0003}$ and $B_0=1.0$.
The apparent power law decay with exponent $1.5$, respectively $4$,  in 
the intermediate, respectively asymptotic, regimes are indicated by straight lines.
}
\end{center}
\end{figure}

\newpage

\begin{figure}
\begin{center}
\epsfig{file=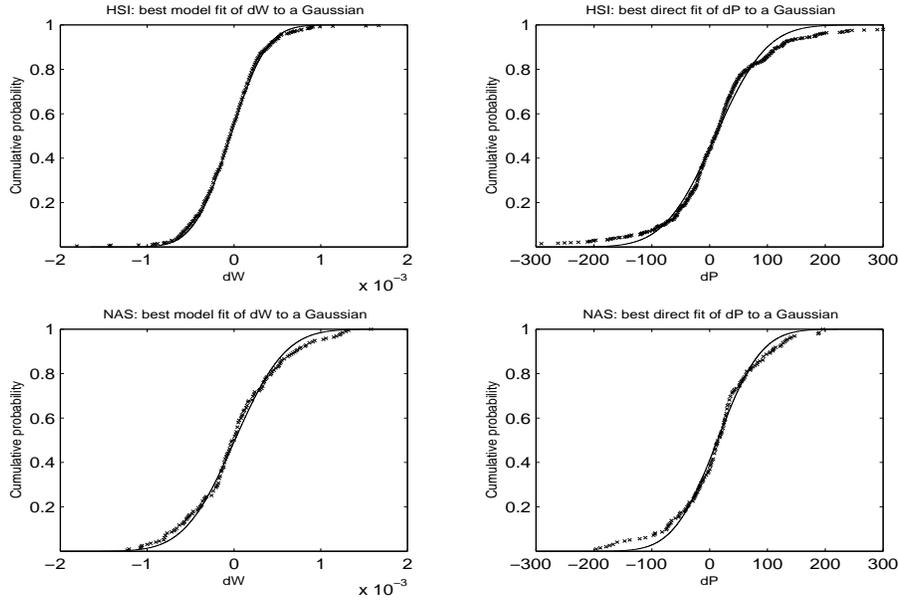,height=8cm,width=12cm}
\caption{\protect\label{Fig4}  Cumulative distributions of the increments $dW$ 
of the best reconstructed ${\hat W}(t)$ given by (\ref{bggnlas})
for the Hang Seng and Nasdaq bubbles
and their fit by a cumulative Gaussian distribution (continuous lines). 
Left (resp. right) panels
correspond to the distribution of the returns of the reconstructed ${\hat W}(t)$ (resp.
of the empirical index prices). Hang Seng index: 
the best fit is with $\alpha=2.5, V=1.1 \cdot 10^{-7}, M=4.23 \cdot 10^{-5},
r=0.00032$ and $F=2267.3$, corresponding to a KS confidence level of $96.3\%$.
This should be compared with the best Gaussian fit to the empirical price returns giving
$V=4879.6, M=10.1$, corresponding to a KS confidence level of $11\%$. 
Nasdaq composite index: $\alpha=2.0, V=2.1 \cdot 10^{-7}, M=-9.29 \cdot 10^{-6}, r=0.00496$
and $F=641.5$, corresponding to a KS confidence level of $85.9\%$. The corresponding
best Gaussian fit to the empirical price gives
$V=3560.3, M=13.51$ corresponding to a KS confidence level of $73\%$.
}
\end{center}
\end{figure}

\newpage

\begin{figure}
\begin{center}
\epsfig{file=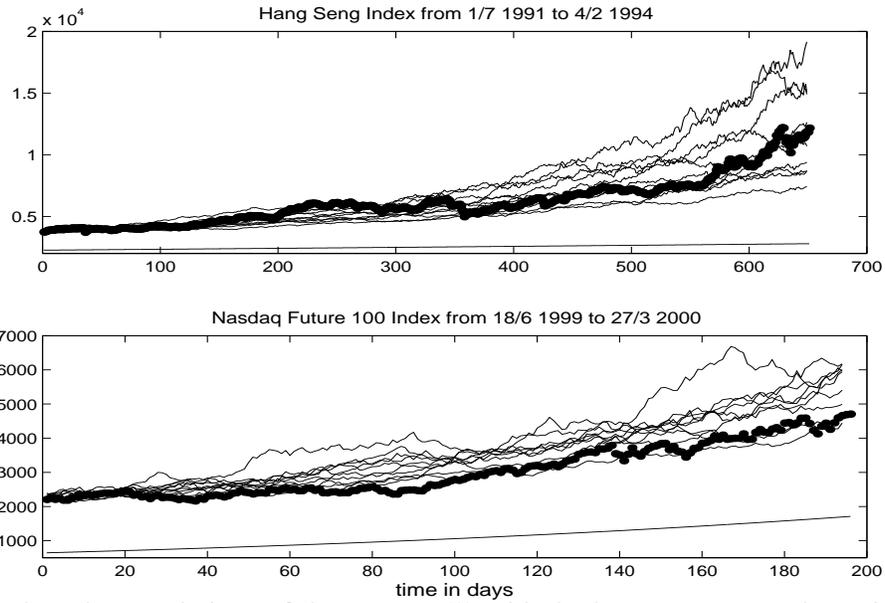,height=8cm,width=12cm}
\caption{\protect\label{Fignadheng}  Ten random time evolutions of 
the process (\ref{fundsum}) with 
the best parameter values given in figure 4 with distinct random realizations of synthetic
random walks ${\hat W}(t)$ for both bubbles and comparison with the empirical prices
shown as the thick lines (one time step corresponds to one trading day). 
The smooth continuous line close to the 
horizontal axis is the fundamental price $F e^{rt}$. One should be very careful
about concluding that the Hang Seng price seems to be mostly a 
bubble growth as the observed price is
much larger and increasing much faster than the fundamental price, since
the quality of the
inversion is almost insensitive to the fundamental value $F$. At present, the model
cannot be used as a reliable calibration of the fundamental value.
}
\end{center}
\end{figure}

\newpage

\end{document}